\documentclass{osa-article}
\journal{osajournal}
\articletype{Research Article}

\newcommand{\Par}{\partial}
\newcommand{\vare}{\varepsilon}

\newcommand{\RE}{\mathrm{Re}}
\newcommand{\IM}{\mathrm{Im}}

\begin{document}

\title{Exceptional points for resonant states on parallel circular dielectric cylinders}

\author{Amgad Abdrabou
 and Ya Yan Lu\authormark{*}}
 
\address{Department of Mathematics, City University of Hong Kong, Kowloon, Hong  Kong} 

\email{\authormark{*}mayylu@cityu.edu.hk}

\begin{abstract} 
Exceptional points (EPs) are special parameter values of a
non-Hermitian eigenvalue problem where eigenfunctions corresponding to
a multiple eigenvalue coalesce.  In optics, EPs are associated with a
number of counter-intuitive wave phenomena, and have potential
applications in lasing, sensing, mode conversion and spontaneous
emission processes. For open photonic structures, resonant states are
complex-frequency solutions of the Maxwell's equations with outgoing
radiation conditions. For open dielectric structures without material
gain or loss, the eigenvalue problem for resonant states can have EPs,
since it is non-Hermitian due to radiation losses. 
For applications in nanophotonics, it is important to
  understand EPs for resonant states on small finite dielectric structures
  consisting of conventional dielectric materials. To achieve this
  objective, we study EPs of resonant states on
finite sets of parallel infinitely-long circular dielectric cylinders
with subwavelength radii. 
 For systems with two, three and four
cylinders, we develop an efficient numerical method for
  computing EPs, present examples for second and third order EPs, and
highlight their topological features. Our work provides insight to
understanding EPs on more complicated photonic structures, and can be
used as a simple platform to explore applications of EPs. 
\end{abstract}

\section{Introduction}

One of the most interesting features of non-Hermitian eigenvalue
problems is the existence of exceptional points (EPs),  
where two or more eigenvalues, as well as their corresponding
eigenfunctions,  coalesce~\cite{Kato,Berry,Heiss04,Moiseyev,Heiss,Miri}.  
For optical systems, EPs exist in eigenvalue problems directly
formulated from Maxwell's equations or indirectly formulated from
scattering operators, and the non-Hermiticity comes from material loss
and/or gain, or radiation loss for open structures. 
Associated with EPs, many interesting wave phenomena have been 
theoretically studied and experimentally
observed~\cite{Miri}. These unique features have found valuable
applications in unidirectional propagation~\cite{Lin},
lasing~\cite{EPLas1,EPLas2,Feng,EPLas3},
sensing~\cite{Wiersig,Chen2017,Hodaei,Langbein,Zhang18}, mode
conversion~\cite{Hassan}, and spontaneous emission
processes~\cite{Pick17}. Many works on EPs and their
  applications are associated with parity-time (${\cal PT}$) symmetric
optical systems with a balanced gain and
loss~\cite{BenderPT,BenderPT2,BenderPT3,PTsym}. In that case, EPs can
be easily found by tuning a single parameter, namely, the amplitude of 
the balanced gain and loss. Since it is not always easy or desirable to
keep a balanced gain and loss in an optical system, it is of significant
interest to explore EPs and their applications in non-${\cal
  PT}$-symmetric optical systems.
EPs can appear in open dielectric systems with a real dielectric
function~\cite{BoZhen,Arslan,Amgad1,Zhou2018,EPCon18}, since 
power can radiate to infinity in open systems, and the
eigenvalue problem for resonant states, directly formulated from
Maxwell's equations with 
an outgoing radiation condition, is non-Hermitian. Currently, there exist a few studies concerning EPs for resonant
  states on periodic dielectric structures sandwiched between two
  homogeneous half-spaces~\cite{BoZhen,Arslan,Amgad1,Zhou2018}. For
  these structures, EPs are relatively 
easy to obtain due to the  
additional degrees of freedom associated with Bloch wavenumbers, they
are related to Dirac points in the band structures of two-dimensional
photonic crystals \cite{BoZhen}, and can be 
traced to some special points in the band structures of uniform slabs
\cite{Amgad1}. In addition, paired EPs on periodic dielectric
structures exhibit interesting topological properties
\cite{Zhou2018}. For non-periodic dielectric structures, EPs can be found
by tuning two or more geometric or material parameters. A multilayered
concentric dielectric cylinder is probably the simplest structure on
which EPs of resonant modes have been found~\cite{EPCon18}.  

For applications in nanophotonics, it is important to have EPs on
structures with a size on the optical wavelength. In addition, it is
also important to realize EPs without tuning the dielectric constant,
since  available dielectric materials are limited in practice. In
this paper, we analyze EPs for a small number of parallel and
infinitely-long dielectric cylinders. The structures are
  chosen for their simplicity. Our objective is to gain insight
  about EPs on small non-periodic structures consisting of
  conventional dielectric materials.  We are interested in EPs of
resonant states with relatively high quality factors and
relatively low resonant frequencies, so that
the radii of the cylinders and the distances between them are smaller
than or about the same as the resonant wavelength. For systems with
two, three and four cylinders, we develop an efficient numerical
method for computing EPs and determine second and third order EPs
where two or three eigenpairs coalesce.

The rest of this paper is organized as follows. In section~\ref{S2}, we
briefly describe the method for computing resonant modes and show an
example where resonant modes exhibit crossing and anti-crossing
behavior with respect to their dependence on a system parameter. In
section~\ref{S3}, we present a method for computing EPs, show second
order EPs for systems with two or more cylinders, and discuss their
topological features.  Third order EPs and related topological
properties are studied in section~\ref{S4}. The paper is concluded with a
brief discussions in section~\ref{S5}. 
 
\section{Resonant modes}\label{S2}

For structures that are invariant along the $z$ axis and the
$E$-polarization, the $z$ component of the electric field, denoted as
$u$, satisfies the following Helmholtz equation 
\begin{equation}
\label{Eq1}
\frac{\Par^2 u}{\Par x^2} +\frac{\Par^2 u}{\Par y^2}+k^2 \vare(x,y)u = 0,
\end{equation} 
where $\vare=\vare(x,y)$ is the dielectric function, $k=\omega/c$ is
the free-space wavenumber, $\omega$ is the angular frequency, and $c$
is the speed of light in vacuum. The time dependence is assumed to be
$\exp(-i\omega t)$.  For the $H$-polarization, the $z$ component of the
  magnetic field, also denoted as $u$, satisfies the following
  slightly different Helmholtz equation
  \begin{equation}
    \label{helmTM}
\frac{\partial}{\partial x} \left( \frac{1}{\varepsilon(x,y)} 
\frac{\partial u}{\partial x}  \right) + 
\frac{\partial}{\partial y} \left( \frac{1}{\varepsilon(x,y)} 
\frac{\partial u}{\partial y}  \right) + k^2 u = 0.
  \end{equation}
A resonance mode (or resonant state) is a non-zero solution of
Eq.~\eqref{Eq1} or Eq.~\eqref{helmTM} satisfying an outgoing radiation condition as
$r:=\sqrt{x^2+y^2} \to \infty$. If the structure is finite (in the
$xy$-plane) and surrounded by air, then $\varepsilon(x,y)= n_0^2$
for sufficiently large $r$,  $n_0\approx 1$ is the refractive index of air,  and the wave field of a resonant mode can be 
expanded as  
\begin{equation}
  \label{farfield}
u({\bf r}) = \sum_{m = -\infty}^\infty \hat{u}_m H_m^{(1)}(kn_0 r) e^{
  i m \theta}, 
\end{equation}
for sufficiently large $r$, where  ${\bf r}=(x,y)$, $\theta$ is the
polar angle, $H_m^{(1)}$ is the Hankel function of first kind and
order $m$, and $\hat{u}_m$ are expansion coefficients.  Since each
term in the right hand side of \eqref{farfield} 
radiates power to infinity and there is no source or incoming wave, a
resonant mode can only exist for a complex $k$, so that it can decay with
time. Under  the assumed time dependence, the imaginary part of $k$
must be negative.  

If the structure consists of $N$ parallel circular cylinders
surrounded by air, where the cylinders have centers ${\bf c}_j$, radii
$R_j$ and dielectric constants $\varepsilon_j$ for $1\le j \le N$, the
classical  multipole method~\cite{Martin,Yasumoto,Felb,Tayeb} is probably the most
efficient method for
solving Eqs.~\eqref{Eq1} and \eqref{helmTM}. Outside the cylinders,  the wave field of a
resonant mode can be expanded as 
\begin{equation}
  \label{multipole}
  u({\bf r}) = \sum_{j=1}^N \sum_{m=-\infty}^\infty b_m^{(j)}
  H_m^{(1)}(k n_0 r_j) e^{ i m \theta_j},
\end{equation}
where $r_j$ and $\theta_j$ are the magnitude and polar angle of 
${\bf r}-{\bf c}_j$, respectively, and  $b_m^{(j)}$ are expansion
coefficients. The multipole method gives  
a homogeneous linear system
\begin{equation}
\label{linearsys}
A(k)\, \mathbf{b} = \mathbf{0},
\end{equation} 
where $\mathbf{b}$ is a column vector for all $b_m^{(j)}$ with $m$
truncated to a finite integer set.  More details about
  Eq.~\eqref{linearsys} are given in Appendix A. Here, we 
  consider Eq.~\eqref{linearsys} as an eigenvalue problem with 
$k$ being the eigenvalue. 
 Since matrix $A$ depends on $k$ nonlinearly, 
it is a nonlinear eigenvalue problem. Notice that both $k$ 
and ${\bf b}$ are unknown, and ${\bf b}$ has to be a nonzero
vector. Since a nonzero ${\bf b}$ is only possible when $A(k)$ is singular,
we can determine $k$ from the condition that $A(k)$ is a singular
matrix. As in~\cite{LuMP}, we solve $k$ from $\lambda_1(A) =0$, where
$\lambda_1(A)$ is the linear eigenvalue of $A$ with the smallest magnitude
and it is regarded as a function of $k$. Once $k$ is
  determined, ${\bf b}$ can be calculated as the  eigenvector
  corresponding to the zero linear eigenvalue of $A$, then the wave
  field outside the cylinders can be evaluated using
  Eq.~\eqref{multipole}, and field inside each cylinder can also be
  evaluated \cite{Felb}. See Appendix A for more details.

\begin{figure}[htbp]
	\centering 
	\includegraphics[scale=0.5]{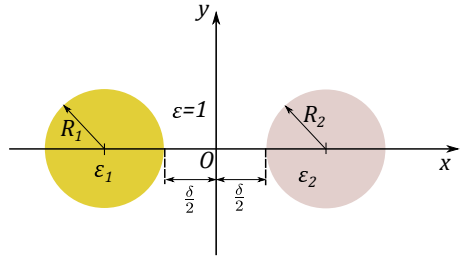}
	\caption{Two circular cylinders parallel to
          the $z$ axis and surrounded by air.}
	\label{Fig1}
\end{figure}

\begin{figure}[htbp]
	\centering 
	\includegraphics[scale=0.5]{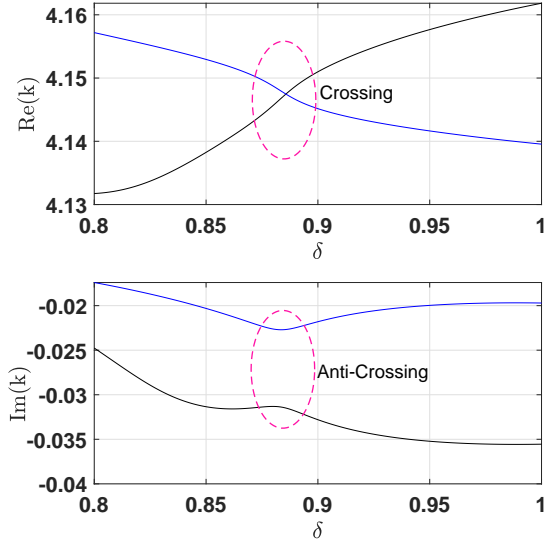}
	\caption{Crossing and anti-crossing in the
real and imaginary parts of $k$ (as functions of distance $\delta$) of
two resonant modes on a system with two cylinders.}
	\label{Fig2n}
\end{figure}

As an example, we consider two cylinders separated by a distance
$\delta$ as shown in Fig.~\ref{Fig1}. 
Assuming the cylinders have dielectric constants $\vare_1 = \vare_2 =
11.56$ and radii $R_1 = 1$ and $R_2 = 1.66$, we calculate resonant
modes of $E$-polarization for a varying spacing $\delta$. 
For simplicity, we take $R_1$ as the characteristic length
  scale to non-dimensionalize other quantities. For example, $R_2$ is
  regarded as $R_2/R_1$, $\delta$ is regarded as $\delta/R_1$, and $k$
  should be interpreted as $kR_1$.
The complex wavenumbers $k$ for
two resonant modes are shown in Fig.~\ref{Fig2n}.  
It can be observed that the real parts of $k$ have a crossing at
$\delta \approx 0.88$, but the imaginary parts of $k$ avoid the
crossing.  In fact, the field patterns of the two modes are quite
similar at the crossing, yet their decay rates are different due to
the anti-crossing in $\IM(k)$.  It is also common for the real parts
of $k$ to exhibit an anti-crossing and the imaginary parts of $k$ to
have a crossing. 


\section{Second order exceptional points}
\label{S3}

A second order EP is a non-Hermitian degeneracy where two
eigenvalues coincide and their eigenfunctions coalesce. In general, it
is necessary to tune two system parameters to find a second order EP. For a set
of circular cylinders, the system parameters include geometric and
material parameters such as the radii and dielectric constants of the
cylinders and the distances between them.  Since available dielectric
materials are limited, it is desired to find EPs by tuning only the
geometric parameters.  For the nonlinear eigenvalue problem given in
Eq.~\eqref{linearsys}, a second order EP occurs at a complex value of $k$
where the matrix $A$ has a double singularity. One approach is to
calculate $k$ from the two conditions $\mathrm{det}(A)= d [ \det(A)]
/dk = 0$ \cite{Heiss3rd}.  If the size of matrix $A$ is not very
small, it is more robust to use the smallest linear eigenvalue of $A$ instead
of its determinant~\cite{LuMP}.  Therefore, we solve $k$ from the
following two conditions 
 \begin{equation}
\label{twolam}
 \lambda_1(A) = \frac{d}{dk} \lambda_1(A) = 0. 
 \end{equation}
A rigorous justification for the second condition above,
    i.e.,  $d\lambda_1(A) /dk = 0$, is given in Appendix
    B. Since $\lambda_1(A)$ is complex, the two conditions in
    \eqref{twolam} give rise to a system of four real equations, and
    it can be used to solve four 
    real unknowns, i.e., the real and imaginary parts of $k$ and
    two system parameters. We use MATLAB function {\tt fsolve} to
    solve the nonlinear system, and it requires a vector function that
    inputs $\mbox{Re}(k)$, $\mbox{Im}(k)$ and two system parameters,
    and outputs real and imaginary parts of $\lambda_1(A)$ and 
    $d\lambda_1(A)/dk$. For a given structure and a given 
    complex $k$, the matrix $A$ can be formed when the
    infinite sum over $m$ in Eq.~\eqref{multipole} is properly truncated,
    $\lambda_1(A)$ can be calculated by MATLAB function {\tt eigs},
    and $d\lambda_1(A)/dk$ can be approximated using a 
    difference formula. Initial guesses for $k$ and the two system
    parameters are needed.

\begin{figure}[htbp]
 	\centering
 	\includegraphics[scale=0.5]{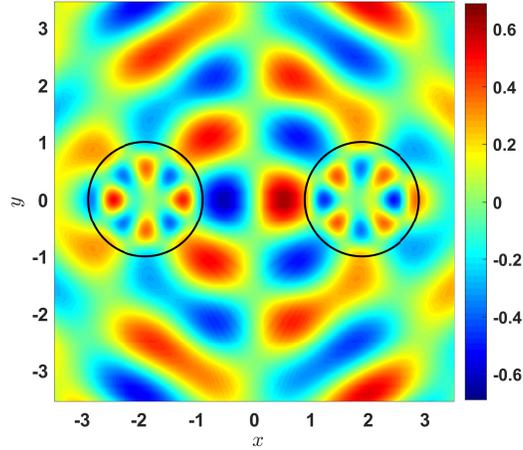}
 	\caption{Electric field (real part of $u$) of a resonant mode at an
          EP on two identical cylinders.} 
 	\label{Fig1ext}
 \end{figure}

 For a single homogeneous circular cylinder, the resonant modes can be
solved analytically. The solutions are well-separated and no EPs can
be found. In a recent work, Kullig {\it et al.} \cite{EPCon18} found
some EPs for an inhomogeneous cylinder with different but concentric
layers. However, their EPs exist at very special values of
dielectric constants. For two or more cylinders, more
geometric parameters are available, and they can be tuned to find EPs.  
As a first example, we consider two identical circular and homogeneous
cylinders separated by a distance $\delta$ and surrounded by air.
Assuming the radius of the two cylinders is $R=1$ (i.e. the radius is
used as the characteristic length scale), the system has only two parameters $\delta$
and $\varepsilon$ (the dielectric constant of the cylinders). We
obtain a second order EP  for the $E$-polarization at 
$\delta = 1.78773$ and $\vare = 6.23690$. The complex wavenumber of
the resonant mode at this EP is 
$k = 3.72476-0.13420i$,  its quality factor is $ Q = -0.5
\mbox{Re}(k)/\mbox{Im}(k) \approx 13.878$, and  the resonant
wavelength is  $2\pi/\mbox{Re}(k) \approx 1.6869$. Since 
$R$ is used as the length scale, the above resonant wavelength should be
interpreted as its ratio with $R$. The 
electric field pattern of the  resonant mode is shown in Fig.~\ref{Fig1ext}. 
The field patterns inside the cylinders resemble two
identical whispering gallery modes with opposite signs. It appears
that the coalescence of these two modes gives rise to a rather strong
field outside the cylinders. Overall, the electric field is odd in $x$
and even in $y$, where $x$ is the horizontal axis passing through the
centers of the two cylinders. For this problem, the two complex
conditions in Eq.~\eqref{twolam} allow us to determine two real parameters
$\delta$ and $\varepsilon$ and the complex wavenumber $k$.

\begin{figure}[h!]
	\centering
	\includegraphics[scale=0.5]{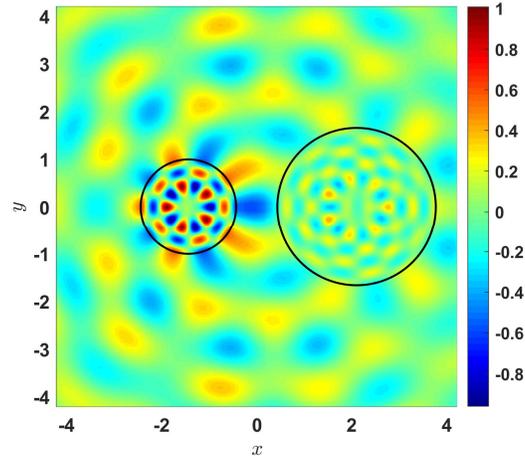}
	\caption{Electric field (real part of $u$) of a resonant mode at
          a second order EP on two cylinders with $\vare_1 = \vare_2 =
          11.56$, $R_1 = 1$, $R_2 = 1.66056$ and
          $\delta=0.88440$. } 
	\label{Fig3n}
\end{figure}

\begin{figure}[h!]
  \centering
  \includegraphics[scale=0.38]{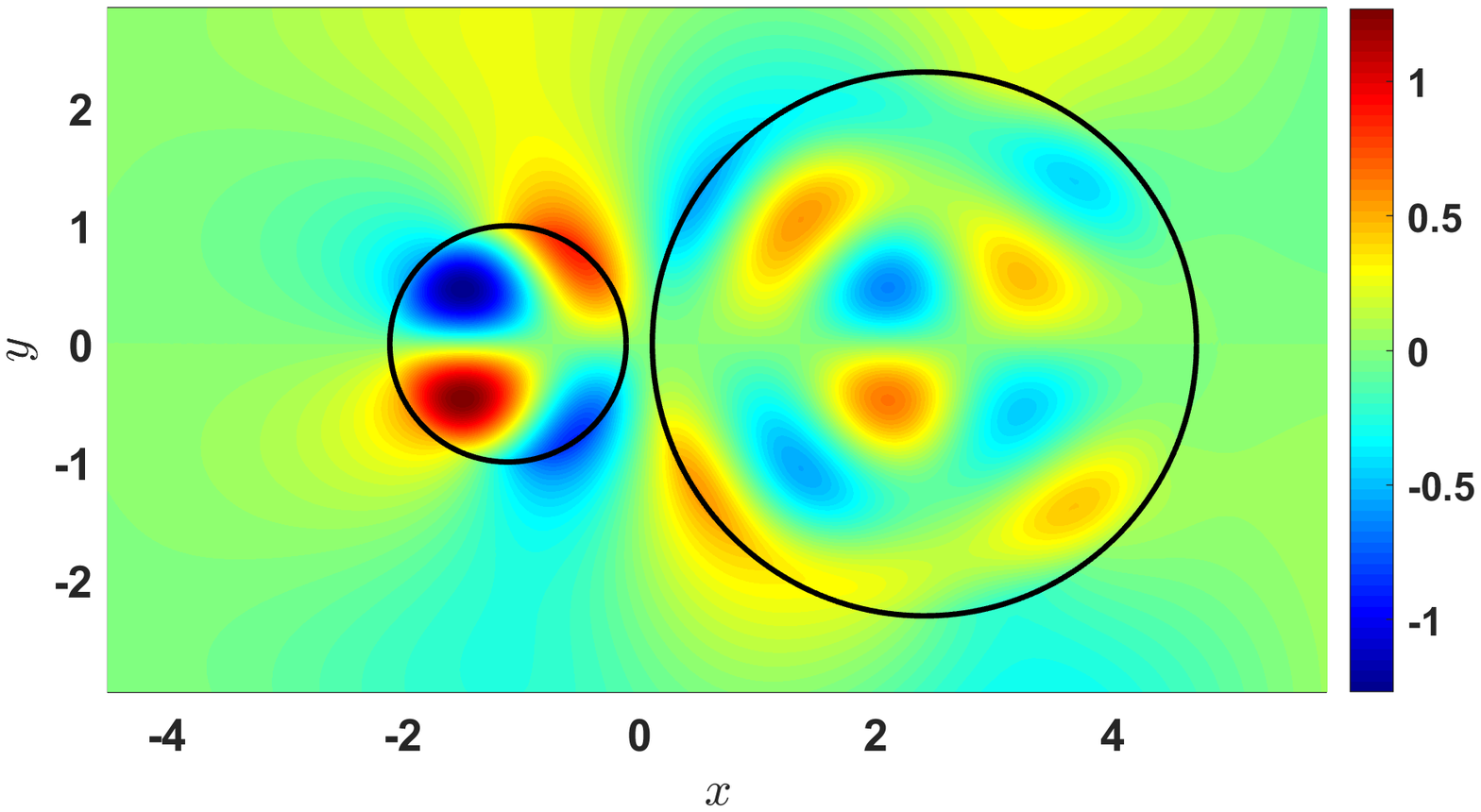}
  \includegraphics[scale=0.38]{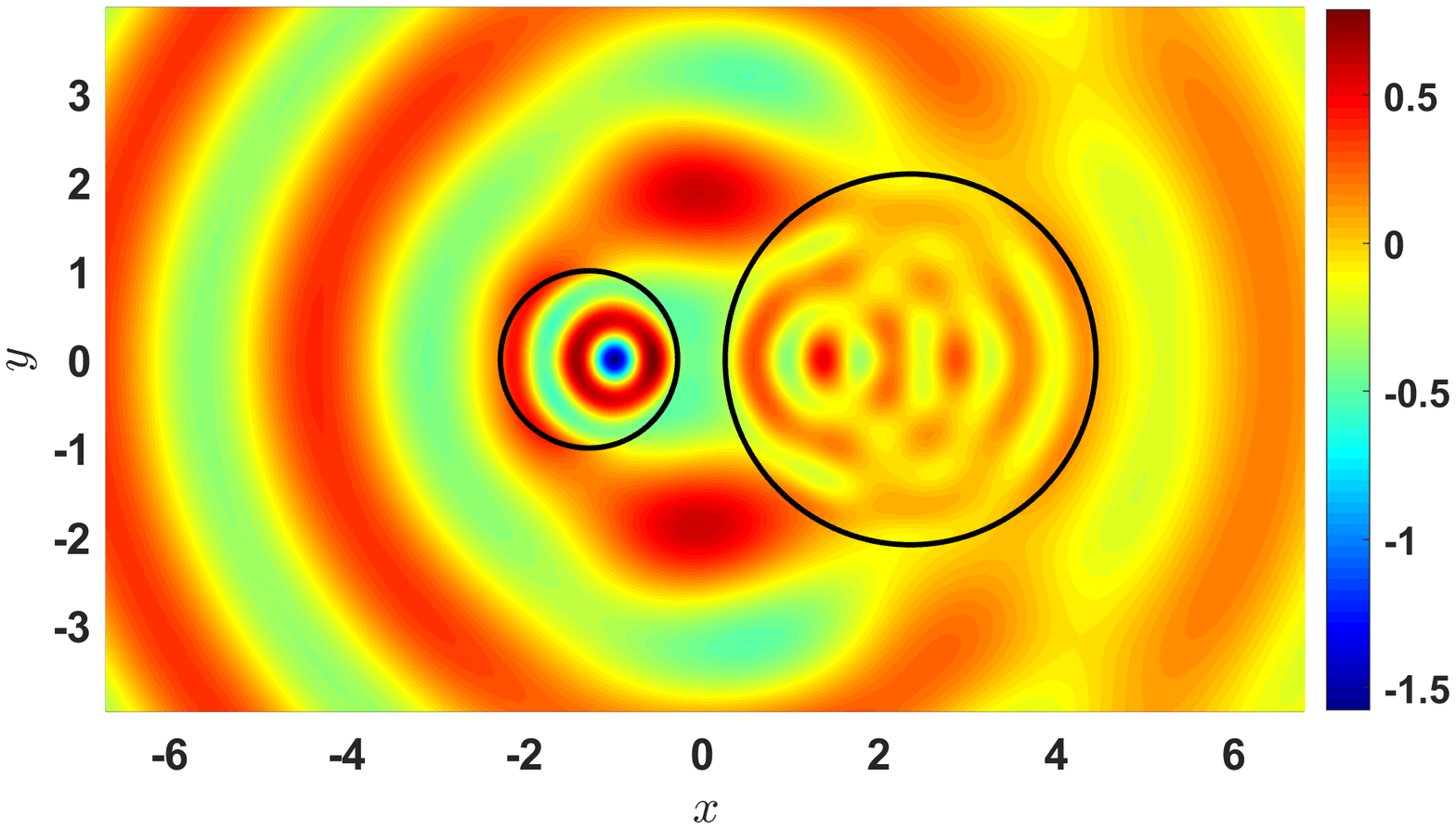}
  \caption{Magnetic field patterns of the resonant modes at two EPs
    for the $H$-polarization. The upper panel is for the first EP at 
$R_2= 2.30294$ and $\delta= 0.22186$,  and the lower panel is for the
second EP at $R_2= 2.09224$ and $\delta=0.53607$.}
  \label{Hpols}
\end{figure}

For real applications, it is unlikely that there is a material with the
precise dielectric constant at which the EP exists. 
For this reason, we revert to a system with two circular cylinders
with different radii as shown in Fig.~\ref{Fig1}. We assume the
dielectric constants of both cylinders are fixed at $\vare_1 = \vare_2
= 11.56$, and the radius of one cylinder is $R_1=1$. The available
geometric parameters are $R_2$ and $\delta$. Solving the two complex
equations in (\ref{twolam}), we obtain a second order EP 
  for the $E$-polarization at 
$R_2  =  1.66056$ and  $\delta= 0.88440$. The complex wavenumber of 
the resonant mode at this EP is 
$k= 4.14671 -0.02706i$, and the $Q$-factor is 76.632. 
The electric field pattern is shown in Fig.~\ref{Fig3n}.
It looks like the field is approximately the superposition of two modes
concentrated in the two cylinders. For the
  $H$-polarization, we found two EPs at 
$R_2= 2.30294$ and 
$\delta= 0.22186$,  and $R_2= 2.09224$ and $\delta=0.53607$,
  respectively. The complex wavenumber and quality factor of the resonant
  mode at the first EP are $k = 1.15695  -0.08219i$ and
  $Q=7.0383$. Those of the resonant mode at the second EP are 
$k = 2.49077  -0.08261i$ and $Q = 15.075$. The magnetic field patterns
of the resonant modes at these two EPs are shown in Fig.~\ref{Hpols}

\begin{figure}[h!]
	\centering 
	\includegraphics[scale=0.4]{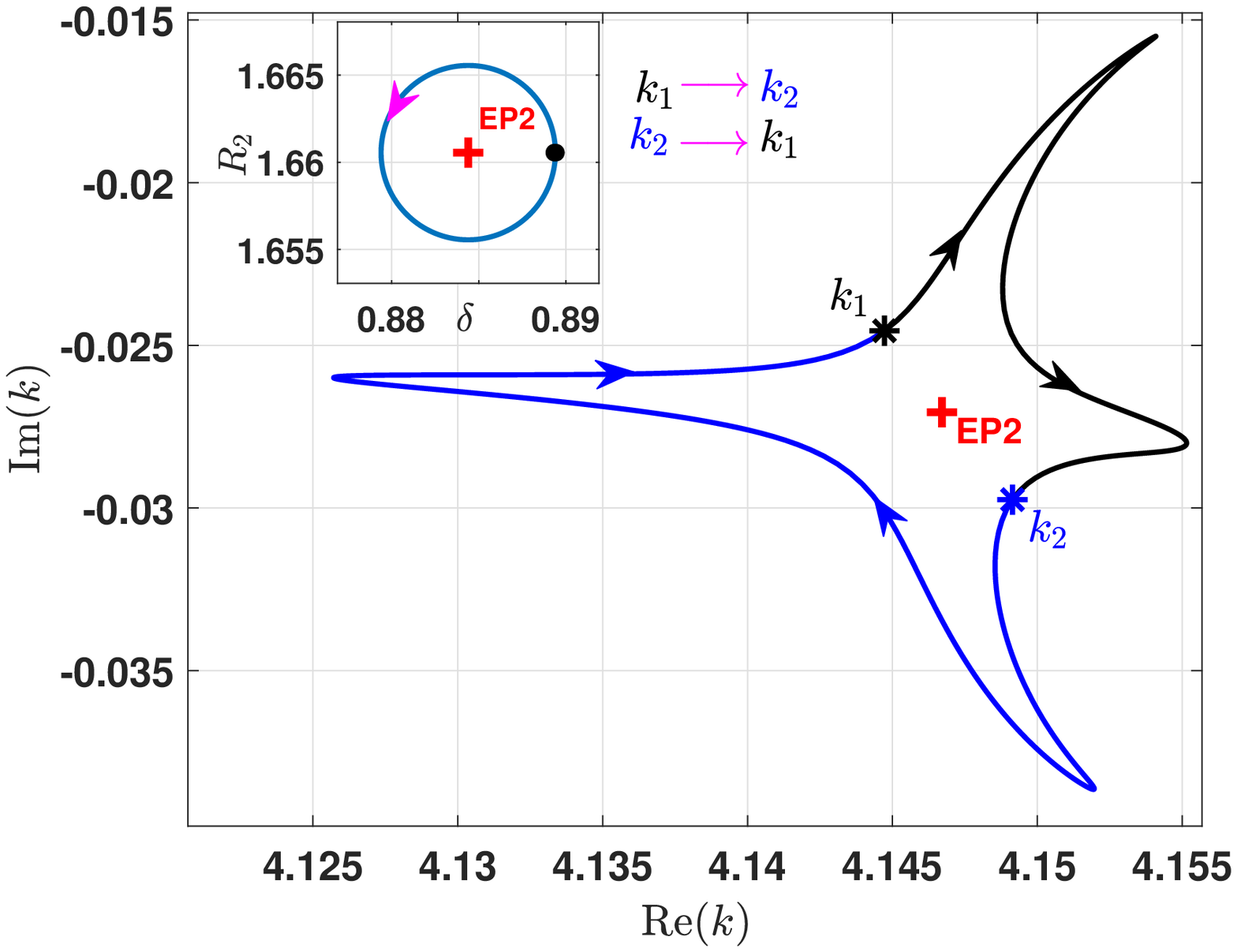}
	\caption{Encircling an EP  in the parameter space,
          $\delta$-$R_2$ plane, showing the switching of the
          eigenvalues $k_1\to k_2$ and $k_2 \to k_1$.} 
	\label{Fig4}
\end{figure}

\begin{figure}[h!]
	\centering
	\includegraphics[scale=0.5]{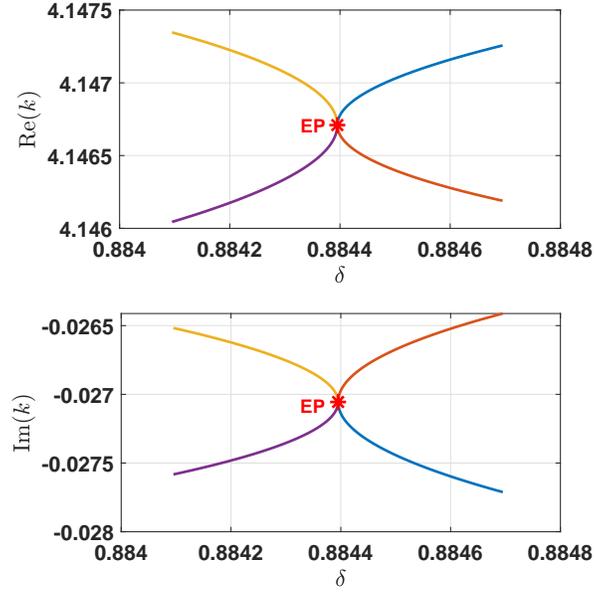}
	\caption{Square-root splitting of the real (top) and imaginary
          (bottom) parts of $k$ versus $\delta$ near a second order EP.}
 \label{Fig5}	
\end{figure}

It is known that an EP can be verified by its topological signature in the
parameter space. Consider the EP for the $E$-polarization
  above, and let $k_1$ and $k_2$ be two eigenvalues emerged from
the eigenvalue at the EP when the parameters move away from
the EP by slightly increasing $\delta$. It is known that encircling an
EP in the parameter space (i.e. the $\delta$-$R_2$ plane) leads the 
eigenvalues $k_1$ and $k_2$ to switch their positions on the
associated Riemann surfaces of $\RE(k)$ and $\IM(k)$. In
Fig.~\ref{Fig4},
 we show the switching  $k_1\to k_2$ and $k_2 \to
k_1$, as $(\delta, R_2)$ moves alone the circle shown in the inset.
The particular parameters that give rise to $k_1$ and $k_2$ correspond
to the dot on the circle. It is well known that a second order EP is a branch point where
solutions manifest square-root splittings in
both $\RE(k)$ and $\IM(k)$. In Appendix B, we show that
  this is true even for nonlinear eigenvalue problems.
In Fig.~\ref{Fig5},
the real and imaginary parts of $k$ are shown as functions of $\delta$
varying around the
EP value 0.88440. The radius $R_2$ is fixed at $1.66056$.

\begin{figure}[htbp]
	\centering 
	\includegraphics[scale=0.5]{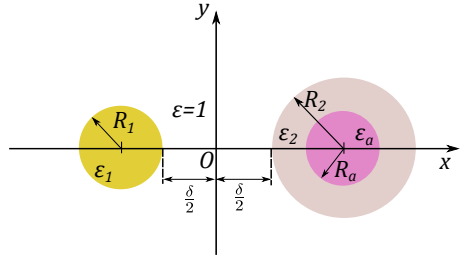}
	\caption{A system of two cylinders of which one has a
          circular core with radius $R_a$ and dielectric constant 
          $\vare_a$.} \label{Fig6}	
\end{figure}

\begin{figure}[h!]
 	\centering
 	\includegraphics[scale=0.5]{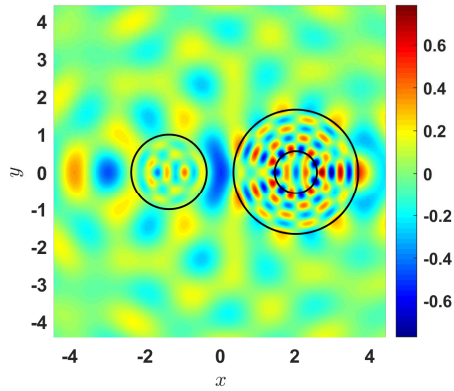}
 	\caption{Electric field (real part of $u$) of a resonant mode
          at an EP on a two-cylinder structure where one cylinder
          has a core.}
 \label{Fig7} 
 \end{figure}

The quality factors of the resonant modes at the above EPs are not very high. Since it is
desirable to have high $Q$ resonances, we consider a 
slightly more complicated system with one inhomogeneous cylinder and
one homogeneous cylinder as shown in Fig.~\ref{Fig6}. 
The  inhomogeneous cylinder consists of a circular core with radius $R_a$ and
dielectric constant $\varepsilon_a$ and a shell with outer radius $R_2$
and dielectric constant $\varepsilon_2$. The other parameters 
$R_1$, $\vare_1$ and $\delta$ are defined as before. Assuming 
$\vare_1 = \vare_2 = 11.56$,  $\vare_a=10.20$, $R_1 = 1$, 
and $R_2 = 1.66$,  the system has two remaining parameters
$\delta$ and $R_a$. An EP for the $E$-polarization is found at 
$\delta= 0.734985$ and $R_a  =  0.561180$. The complex wavenumber and 
quality factor of the resonant mode at this EP is 
$k  = 4.16669 -0.01824i$ and $Q = 114.20$. Its electric field  is
shown in Fig.~\ref{Fig7}. 
 
\begin{figure}[h!]
	\centering 
	\includegraphics[scale=0.65]{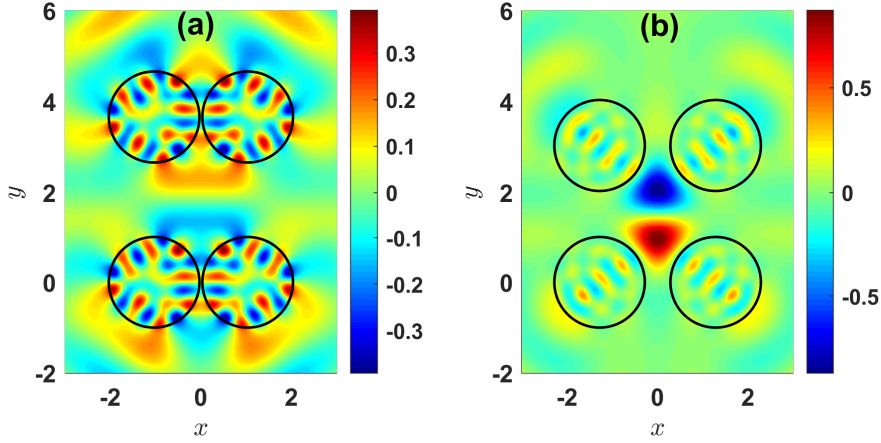}
	\caption{Electric fields (real part of $u$) of resonant modes
          at two EPs on a 
          system of four parallel identical cylinders.}
	\label{Fig8} 
\end{figure}

The next example consists of four identical cylinders with their
centers located at the corners of a rectangle with sides parallel to
the $x$ and $y$ axes. Assuming the radius and
the dielectric constant of the cylinders are $R=1$ and $\vare_1 =
11.56$, the system still has two free parameters $\delta_x$ and
$\delta_y$, i.e, the horizontal and vertical distance between two cylinders. Using
the same method, we obtain two second order EPs for the
$E$-polarization. The results are listed in Table~1  
below. 
\begin{table}
  \centering
  \begin{tabular}[h]{c|c|c} \hline
    EP  & (a) & (b) \\ \hline
    $\delta_x$ & 0.06453 & 0.56679 \\ \hline
    $\delta_y$ & 1.65891 &  1.03352 \\ \hline
    $\mbox{Re}(k)$ & 3.21529  & 3.58809 \\ \hline
    $\mbox{Im}(k)$ & -0.01081 & -0.01517\\  \hline
     $Q$-factor &  148.73&  118.25 \\ \hline
  \end{tabular}
  \caption{Second order EPs for a system of four identical cylinders 
    with radius $R=1$, dielectric constant $\vare=11.56$,
    horizontal distance $\delta_x$ and  vertical distance $\delta_y$.}
  \label{tab111}
\end{table}
The electric field patterns of the resonant modes at these two EPs, labeled as (a)
and (b), are shown in Fig.~\ref{Fig8}. 
Exceptional point (a) is obtained when the
horizontal distance is very small compared with the radius. Strong
electric field appears inside the cylinders. Exceptional point (b) has a strong
field at the center of the structure and outside four cylinders. 
The quality factors of EPs (a) and (b) are  $148.73$ and $118.25$, respectively.


For the above examples, the resonant wavelength, given by
$(2\pi)/\mbox{Re}(k)$, is close to the diameter of the
reference cylinder. The size of the structure in the $xy$ plane is about 3 to
4 wavelengths. Exceptional points at which the resonant
  modes have  higher quality factors can be obtained if one
considers higher resonant frequencies (smaller resonant wavelengths)
or increases the dielectric constants of the cylinders. 


\section{Third-order exceptional points}
\label{S4}
 
Higher order EPs have extra advantages in some applications
\cite{Hodaei,Pick17}. For our problem formulated as Eq.~\eqref{linearsys}, a
third order EP occurs only when the matrix $A$ has a triple
singularity. As shown in Appendix B, we can find third
order EPs by solving the following equations  
\begin{equation}
\label{threelam}
\lambda_1(A) = \frac{d}{dk} \lambda_1(A) =\frac{d^2}{dk^2}
\lambda_1(A)  = 0.  
\end{equation} 
In general, a third order EP can only be found by tuning four system 
parameters. The three complex equations above are equivalent to six 
real equations, and they can only be satisfied if there are six real 
unknowns. Since the complex $k$ corresponds to two real unknowns, 
four additional real parameters are needed. 

\begin{figure}[h!]
	\centering 
	\includegraphics[scale=0.5]{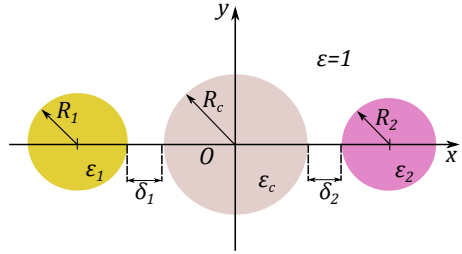}
	\caption{Three parallel dielectric cylinders centered on the
          $x$-axis.}
\label{Fig9}
\end{figure}

\begin{figure}[h!]
	\centering 
	\includegraphics[scale=0.5]{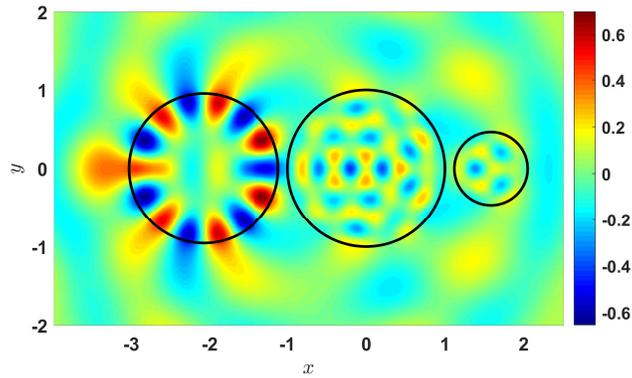}
	\caption{Electric field (real part of $u$) at a third order EP on 
          a system of three cylinders.}
\label{Fig10}	
\end{figure}

In the following, we present examples of third order EPs in a system with three
collinear parallel cylinders as shown in Fig.~\ref{Fig9}.
The centers of the three cylinder are all located on the
$x$ axis. The radius and dielectric constant of the cylinder at the
center is denoted as $R_c$ and $\vare_c$, and the distances between the
nearby cylinder are $\delta_1$ and $\delta_2$. We first assume 
$R_c=1$, $\vare_c=11.56$, $\delta_1 = \delta_2 = 0.12$, and consider 
 $R_1$, $R_2$, $\vare_1$ and $\vare_2$ as free parameters. Solving
 Eq.~\eqref{threelam} for the $E$-polarization, we obtain a third order 
EP for 
$R_1=0.95461$,  $R_2=0.46557$,  $\vare_1=4.44741$, $\vare_2=13.55975$,
 and $k = 4.82031- 0.03264i$.  
The quality factor of the resonant mode at this EP is $Q = 73.831$. 
The field pattern of the resonant mode is shown in Fig.~\ref{Fig10}.
A strong field around the first cylinder can be observed, and it
resembles a whispering gallery mode of azimuthal order $m=7$. 

\begin{figure}[h!]
	\centering 
	\includegraphics[scale=0.45]{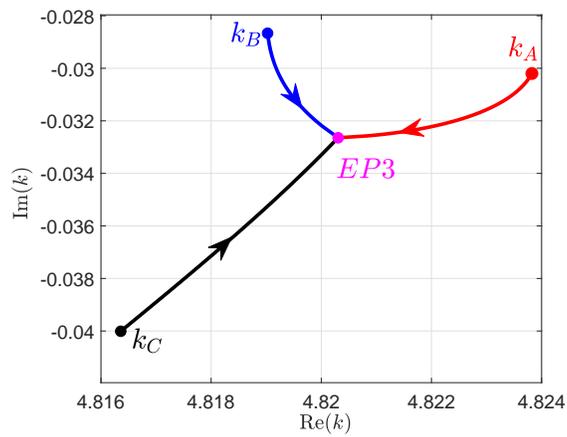}
	\caption{The coalescence of three eigenvalues $k_A$, $k_B$ and
          $k_C$ as parameter $\delta_1$ is decreased from $0.125$
          to $0.12$.}
 \label{Fig11}	 
\end{figure}

\begin{figure}[h!]
	\centering 
	\includegraphics[scale=0.45]{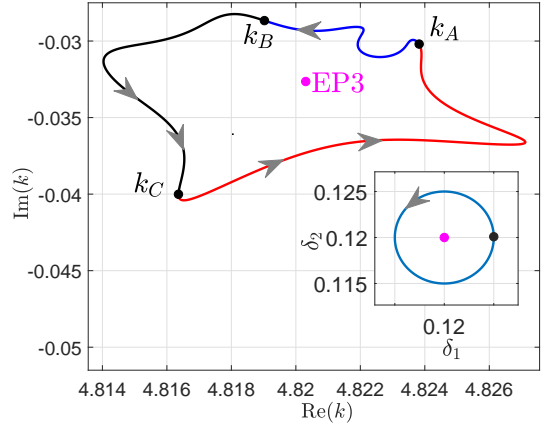}
	\caption{Encircling a third order EP3 in the
          $\delta_1$-$\delta_2$ plane. The arrows in the complex $k$ 
          plane follow the arrow in the inset.}
 \label{Fig12}	 
\end{figure}

 To verify the EP, we first perturb the system by setting $\delta_1=0.125$
 away from the EP condition, and keep $R_1$, $R_2$ and $\delta_2$
 fixed. Three distinct eigenvalues appear near the complex $k$ of 
 the third order EP, and they are denoted as $k_A$, $k_B$ and  
 $k_C$. As $\delta_1$ is continuously decreased to $0.12$, the three
 eigenvalues coalesce at the EP as shown in Fig.~\ref{Fig11}.
Next, we encircle the EP in the parameter space,
$\delta_1$-$\delta_2$ plane, in the counterclockwise
  direction as shown in the inset of Fig.~\ref{Fig12}. 
The circle in the $\delta_1$-$\delta_2$ plane includes the point 
$(\delta_1, \delta_2) =(0.125, 0.12)$  (the black dot in the
inset) that gives rise to $k_A$, $k_B$
and $k_C$. As the parameters $\delta_1$ and $\delta_2$ move along the
circle, the three eigenvalues follow the paths shown in Fig.~\ref{Fig12}. 
After one round in the parameter space,  the eigenvalues switch  as
$k_A\to k_B$,\, $k_B \to k_C$,\, and  $k_C \to k_A$. 
If we encircle the EP in the clockwise direction, then
  the eigenvalues switch as 
$k_A\to k_C$,\, $k_B \to k_A$,\, and  $k_C \to k_B$. This is a
well-known topological signature of EPs.

For this system of three cylinders, it is also possible to find third
order EPs without tuning the dielectric constants $\vare_1$, $\vare_2$ or
$\vare_c$. As an example,  we let  
$\vare_1= 4.45$, $\vare_2  = 13.50$,  $\vare_c=11.56$ and $R_c = 1$, and search
 the four geometric parameters $R_1$, $R_2$, $\delta_1$ and $\delta_2$
 for third order EPs. One third order EP is found at
$R_1 = 0.95426$, $R_2=0.46659$, 
$\delta_1 = 0.11956$ and $\delta_2 = 0.11906$. The complex wavenumber
of the resonant mode at this EP is 
$k = 4.82056 - 0.03269i$. Its field pattern is nearly
identical to the one obtained for fixed $\delta_1=\delta_2=0.12$. 

\section{Conclusion}
\label{S5}

Motivated by potential applications in nanophotonics, we analyzed EPs
for resonant states on simple structures consisting of a few parallel circular
dielectric cylinders surrounded by air. 
 Second and third order EPs are
determined for a few structures involving two to four cylinders.
They are further validated by analyzing the switching of eigenvalues,
a topological signature of the EPs, when pairs of parameters encircle the
EPs in parameter spaces. Importantly,
EPs can be achieved for structures with given dielectric constants by
tuning only the geometric parameters. It is found that
the quality factors of the resonant modes at the EPs can be more than
100 when the structure consists of ordinary dielectric materials and
the overall size of the structure is restricted to about 3 to 4
wavelengths. Depending on the particular configuration, the wave field
of a resonant mode at an EP can be concentrated in or outside the cylinders. For
simplicity, we studied only simple two-dimensional structures with
infinitely-long and circular cylinders. It is anticipated that if
cylinders with more complicated shapes are allowed, the resonant modes at the EPs
may have more desirable field distributions.  It is known
  that on carefully designed structures with restrictions on 
  dielectric constants and size as in this paper, some resonant modes
  unrelated to EPs  can  have very large quality factors
  \cite{highQ}. It is important  to find out whether
resonant modes at EPs on similar structures can have significantly
larger quality factors.
For practical applications, the cylinders of finite height must be studied. These
studies may benefit from the recently developed vertical mode
expansion method (VMEM) for scattering problems involving multiple
non-circular cylinders of finite height \cite{Shi:15,Shi:16,Shi:18}

\section*{Appendix A}
For a system with $N$ circular cylinders surrounded by a homogeneous
medium of refractive index $n_0$, the matrix $A$ in Eq.~\eqref{linearsys}
is an $N \times N$ block matrix, ${\bf b}$ is a vector with  $N$ blocks, and
they are
\[
A = \left[ \begin{matrix} I & - S_1 T_{12} & - S_1 T_{13} &\cdots \cr 
- S_2 T_{21} & I & - S_2 T_{23} &\cdots \cr 
- S_3 T_{31} & - S_3 T_{32} &I &\cdots \cr 
\vdots &\vdots &\vdots &\ddots 
\end{matrix}
\right], \quad 
{\bf b} = \left[ \begin{matrix} 
{\bf b}_1 \cr 
{\bf b}_2 \cr {\bf b}_3 \cr \vdots 
\end{matrix} \right],
\]
where ${\bf b}_j$ is a column vector of $b_m^{(j)}$ for all $m$, 
$S_j$ is a diagonal matrix, and $I$ is the identity matrix. 
Assuming the center, radius and refractive index of the $j$th cylinder
are  ${\bf c}_j$, $R_j$ and $n_j$,  respectively,  then the $(m,q)$ entry
of matrix $T_{jp}$ is 
$(T_{jp})_{mq} = H_{m-q}^{(1)}(k n_0 r_j^p) \exp[i(q-m)\theta_j^p]$, 
where $(r_j^p , \theta_j^p)$ are the polar coordinates of 
${\bf c}_p - {\bf c}_j$. For the $E$ polarization, the $(m,m)$ entry 
of $S_j$ is 
\[
(S_j)_{mm} = \frac{n_j J_m(\xi) J'_m(\eta) - n_0 J_m(\eta) 
J'_m(\xi)} {-n_j H_m^{(1)}(\xi) J'_m(\eta) + n_0 
J_m(\eta) {H_m^{(1)}}'(\xi)}, 
\]
where $\xi  = k n_0 R_j$ and 
$\eta = k n_j R_j$. For the $H$ polarization, the  formula for
$(S_j)_{mm}$ can be obtained by swapping $n_0$ and $n_j$.
If we choose a positive integer $m_*$ and truncate $m$ to $-m_* \le  m \le  m_*$, 
then ${\bf b}_l$ is a vector of length $M=2m_* + 1$, $T_{jp}$ and
$S_j$ are $M \times M$ matrices, and $A$ is an $(MN)\times(MN)$
matrix. More details can be found in \cite{Felb}. 

\section*{Appendix B}

To justify Eqs.\eqref{twolam} and \eqref{threelam}, we need a
perturbation theory about how eigenvalues depend on parameters near 
an EP for nonlinear eigenvalue problems. For linear eigenvalue
problems, this is the well-known  power-$1/n$ splitting for $n$-th
order EPs \cite{Pick17}. In the following, we clarify the results for
second and third order EPs in nonlinear eigenvalue problems. 

Consider the following nonlinear eigenvalue problem
\begin{equation}
  \label{Bnev}
N(\beta, \lambda) v = 0,  
\end{equation}
where $N$ is a square matrix, $\beta$ is a parameter,  $\lambda$ is the
eigenvalue, and $v$ is the eigenvector (a non-zero
vector). There is also a left eigenvector $w$ such that $w^T N(\beta,
\lambda) = 0$. Clearly, $\lambda$, $v$ and $w$ all depend on
$\beta$.

For a second order EP, we have two eigenvalues $\lambda_j(\beta)$ ($j=1,2$)
and eigenvectors $v_j(\beta)$ and $w_j(\beta)$ ($j=1, 2$), and they
satisfy 
$\lambda_j(\beta) \to \lambda_*$, 
$v_j(\beta) \to v_*$,  $w_j(\beta) \to w_*$ as $\beta\to \beta_*$,
where $\beta_*$ is the parameter value for the EP. 
For $\beta$ near $\beta_*$, we have 
\[
N(\beta, \lambda_1) = 
N(\beta, \lambda_2) + (\lambda_1-\lambda_2) N'(\beta,
\lambda_2) + \frac{1}{2} (\lambda_1 - \lambda_2)^2 N''(\beta, \lambda_2) + ... 
\]
where $N'$ and $N''$ denote first and second order partial derivatives of $N$
with respect to $\lambda$.  Since $w_1^T N(\beta,\lambda_1) = 0$, we have
\[
0 = \frac{1}{\lambda_1-\lambda_2} w_1^T N(\beta, \lambda_1) v_2
= 
w_1^T \left[ N'(\beta, \lambda_2) + \frac{1}{2} (\lambda_1 - \lambda_2) 
N''(\beta, \lambda_2) + ... \right] v_2.
\]
Taking the limit as $\beta \to \beta_*$, we obtain the following
self-orthogonality condition
\begin{equation}
  \label{Bself}
w_*^T N'_* v_* = 0,
\end{equation}
where $N'_* = N'(\beta_*, \lambda_*)$. 

In the following, we denote $\lambda_1$ and $\lambda_2$ simply as
$\lambda$, and expand the eigenvalues, eigenvectors and the matrix $N$ in a power
series of $\sqrt{\delta}$, where $\delta=\beta-\beta_*$. The use of
$\sqrt{\delta}$ is necessary, since otherwise the perturbation theory
will become inconsistent. We have 
\begin{eqnarray}
\label{Blam2}
&& \lambda(\beta) = \lambda_* + C_1 \sqrt{\delta} + C_2 \delta + ...
   \\
&& v(\beta) = v_* + v^{(1)} \sqrt{\delta} + v^{(2)} \delta + ... \\
&& w(\beta) = w_* + w^{(1)} \sqrt{\delta} + w^{(2)} \delta + ... \\
&& N(\beta,\lambda) = N_* + C_1 N'_* \sqrt{\delta} + 
(\partial_\beta N_* + C_2 N'_* + 0.5 C_1^2 N''_*) \delta + ...
\end{eqnarray}
where $N_*$ denotes $N(\beta_*, \lambda_*)$, etc. Inserting the above
to $N(\beta, \lambda) v(\beta) =0$, we have  $N_* v_*=0$ at $O(1)$ and
$N_* v^{(1)} + C_1 N'_*  v_* = 0$ at $O(\sqrt{\delta})$.  Due to the
self-orthogonality, the equation  
\begin{equation}
  \label{By1}
N_*  y_1 = - N'_* v_*  
\end{equation}
has a solution. Thus,  $v^{(1)} = C_1  y_1$. At $O(\delta)$, we have 
\[
N_* v^{(2)} + \partial_\beta N_*  v_* + C_2  N'_* v_*   + 0.5 C_1^2 
N''_*  v_* +  C_1   N'_* v^{(1)} = 0.
\]
Multiplying the above by $w_*^T$, we get 
\begin{equation}
  \label{BC1EP2}
C_1^2  = - \frac{w_*^T \partial_\beta N_* v_*}{ w_*^T (N'_* y_1  + 0.5 
  N''_*  v_* )}.  
\end{equation}
The two solutions of $C_1$ (with $\pm$ signs) correspond to the two
eigenvalues. 

For a third order EP, we have three eigenvalues $\lambda_j(\beta) \to
\lambda_*$, three eigenvectors $v_j(\beta) \to v_*$, and three left eigenvectors 
$w_j(\beta) \to w_*$,  as $\beta \to \beta_*$ for $j=1$, 2, 3,  where $\beta_*$
corresponds to the EP. For $\delta = \beta - \beta_*$, we assume 
\begin{eqnarray}
\label{Blam3}
&& \lambda (\beta) = \lambda_* + C_1 \delta^{1/3}  + C_2 \delta^{2/3}
   + C_3 \delta    +  ...   \\
&&  v(\beta) = v_* + 
\delta^{1/3} v^{(1)} + \delta^{2/3} v^{(2)}  +
   \delta v^{(3)} +  ...   \\
&&  w(\beta) = w_* + 
\delta^{1/3} w^{(1)} + \delta^{2/3} w^{(2)}  +
   \delta w^{(3)} +  ...   
\end{eqnarray}
and then 
\begin{eqnarray*}
&&  N(\beta, \lambda) = N_* +  C_1  N'_* \delta^{1/3} + 
( C_2  N'_* + 0.5 C_1^2 N''_*) \delta^{2/3} + \cr
&& \hspace{1.3cm}  + (\partial_\beta N_* + C_3 N'_* + C_1 C_2 N''_* + \frac{1}{6}
   C_1^3 N'''_* ) \delta + ...   
\end{eqnarray*}
As before, we have the self-orthogonality condition \eqref{Bself}
that guarantees the existence of $y_1$ satisfying
Eq.~\eqref{By1}. Under proper assumptions on the three eigenpairs, we
can establish the following condition 
\begin{equation}
  \label{selfo2}
w_*^T \left( N'_* y_1 + \frac{1}{2} N''_* v_* \right)  = 0,
\end{equation}
then there is a vector $y_2$ satisfying
\begin{equation}
  \label{By2}
N_* y_2 = - 
N'_*  y_1 - \frac{1}{2} N''_*  v_*.
\end{equation}
The perturbation theory gives $v^{(1)} = C_1 y_1$,  $v^{(2)} = C_1^2 y_2 + C_2 y_1$,
and finally 
\begin{equation}
\label{BC1EP3}
C_1^3 = - \frac{ w_*^T \partial_\beta N_* v_*}
{ w_*^T \left[  N'_*  y_2 + (1/2) N''_* y_1 
+ (1/6) N'''_* v_* \right] }.
\end{equation}
The three solutions of Eq.~\eqref{BC1EP3} correspond to the three
eigenvalues. 

From Eqs.~\eqref{Blam2} and \eqref{Blam3}, it is clear that we can
write $\sqrt{\delta}$ or $\delta^{1/3}$ as power series of
$\lambda-\lambda_*$ for second and third order EPs,
respectively. Therefore, for a $n$th order EP, we have
\[
\beta-\beta_* = D_1 (\lambda-\lambda_*)^n + ...
\]
for some constant $D_1$. 

If a nonlinear eigenvalue problem of matrix $A$, depending on 
eigenvalue $\lambda$ and some parameters, has an EP with eigenvalue 
$\lambda_*$, eigenvector $v_*$ and left eigenvector $w_*$, 
we can fix the parameters in $A$ at their EP values and introduce a
new parameter $\beta$ and a matrix 
\[
N(\beta, \lambda) = A(\lambda)  - \beta I,
\]
where  
$I$ is the identity matrix. This gives rise to a nonlinear eigenvalue problem for
$N(\beta,\lambda)$ and it has an EP at $\beta=0$ with eigenvalue
$\lambda_*$, eigenvector $v_*$ and left eigenvector $w_*$. Following
the theory developed earlier, we have 
\[
\beta - 0 = D_1 (\lambda-\lambda_*)^n + ...
\]
where $n$ is the order of the EP. Therefore, regarding $\beta$ as a
function of $\lambda$, we have $\beta(\lambda_*) =0$,
$\beta'(\lambda_*) =0$ at a second order EP, where $\beta'$ denotes
the derivative with respect to $\lambda$. For a third order EP, in
addition to these two conditions, we have  
$\beta''(\lambda_*) = 0$. 

In the main sections of this paper, the nonlinear eigenvalue problem
is for matrix $A$ in Eq.~\eqref{linearsys} and $k$ is the eigenvalue,
the extra ``parameter'' introduced is the smallest (in magnitude)  linear eigenvalue
of matrix $A$, denoted as $\lambda_1$. Therefore, Eqs.~\eqref{twolam}
and \eqref{threelam} are obtained when we replace $\lambda$ and
$\beta$ of this Appendix by $k$ and $\lambda_1$, respectively. 

\section*{Funding}
The Research Grants Council of Hong Kong Special Administrative
Region, China (Grant No. CityU 11305518). 

\bibliography{amgad3}

\end{document}